\documentclass[conference]{IEEEtran}
\pdfoutput=1
\usepackage[hyphens]{url}
\usepackage{hyperref}
\hypersetup{breaklinks=true, colorlinks,allcolors=blue}
\usepackage[backend=biber, style=ieee]{biblatex}
\usepackage[top=.75in,left=.75in,right=.75in]{geometry}

\usepackage{amsmath,amssymb,amsfonts}
\usepackage{algorithmic}
\usepackage{graphicx}
\usepackage{textcomp}
\usepackage{xcolor}
\usepackage{balance}
\usepackage{booktabs}
\usepackage{tikz}
\usepackage{array}
\usepackage{float}
\usepackage{lipsum}
\usepackage{soul}

\graphicspath{{./images/}}

\usepackage{balance}

\begin{document}

\date{}

\title{CEKER: A Generalizable LLM Framework for Literature Analysis with a Case Study in Unikernel Security}

\author{\IEEEauthorblockN{Alex Wollman}
\IEEEauthorblockA{\textit{The Beacom College of Computer and Cyber Sciences} \\
\textit{Dakota State University}\\
Madison, SD, USA \\
0009-0000-6260-2750}
\and
\IEEEauthorblockN{John Hastings}
\IEEEauthorblockA{\textit{The Beacom College of Computer and Cyber Sciences} \\
\textit{Dakota State University}\\
Madison, SD, USA \\
0000-0003-0871-3622}
}

\maketitle

\begin{abstract} \label{Abstract}
Literature reviews are a critical component of formulating and justifying new research, but are a manual and often time-consuming process. This research introduces a novel, generalizable approach to literature analysis called CEKER which uses a three-step process to streamline the collection of literature, the extraction of key insights, and the summarized analysis of key trends and gaps. Leveraging Large Language Models (LLMs), this methodology represents a significant shift from traditional manual literature reviews, offering a scalable, flexible, and repeatable approach that can be applied across diverse research domains.

A case study on unikernel security illustrates CEKER's ability to generate novel insights validated against previous manual methods. CEKER’s analysis highlighted reduced attack surface as the most prominent theme. Key security gaps included the absence of Address Space Layout Randomization, missing debugging tools, and limited entropy generation, all of which represent important challenges to unikernel security. The study also revealed a reliance on hypervisors as a potential attack vector and emphasized the need for dynamic security adjustments to address real-time threats.
\end{abstract}


\section{Introduction} \label{Introduction}

As academic research rapidly expands in volume and scope across many fields of study~\cite{thelwall2022scopus}, thorough and efficient systematic literature reviews have become increasingly important~\cite{haddaway2015making}. Literature reviews serve as the foundation for identifying primary themes, research gaps, and emerging trends, thereby informing future research and innovation. However, the traditional approach to literature reviews is quickly being replaced by more modern approaches in order to reduce manual, time-consuming efforts which are prone to inconsistencies in the selection, analysis and synthesis of information from large bodies of academic papers~\cite{bolanos2024artificial,ofori2024towards,wittenborg2024swarm,ali2024automatedliteraturereviewusing,joos2024cuttingclutterpotentialllms}.

This research addresses these challenges by introducing a novel, generalizable approach to literature analysis. The method is designed to be efficient, flexible, and applicable across any literature-backed research domain. Unlike conventional methods that require manual collection and review of literature, this approach supports both the reuse of existing literature collections and the generation of new collections using either traditional or AI-based search strategies. By utilizing large language models (LLMs) to extract key features, analyze trends, and identify research gaps, this approach streamlines the literature review process while also maintaining rigor and reproducibility.

This approach, named CEKER (pronounced \textit{seeker}), 
is a three-step literature review approach that can be generally applied to any research domain as follows:
\begin{enumerate}
\item \textbf{C}\textit{ollect corpus}: Reuse a previously collected corpus of literature if available or collect new literature using traditional search methods or AI tools.
\item \textbf{E}\textit{xtract} \textbf{K}\textit{ey features}: Using an LLM, analyze each paper individually to extract themes related to the topic of interest. Collate these results.
\item \textit{Analyze and} \textbf{E}\textit{valuate} \textbf{R}\textit{esults}: Using an LLM, analyze the results from Step 2 to produce a summary analysis which provides insights into trends, correlations, and gaps, which can be used to guide future research.
\end{enumerate}

To demonstrate the applicability of CEKER, the research applies this method to the domain of unikernel security, a niche but important field within operating system research. The results not only demonstrate the general utility of CEKER, but also provide a comprehensive analysis of unikernel-related literature, identifying key security trends, research gaps, and areas for future exploration. We compare these findings against those from a previous study of unikernel security \cite{wollman2024survey} to highlight new insights that emerge from the refined analysis process.

The remainder of this paper is organized as follows: Section \ref{GeneralApproach} details the general steps in CEKER. Section \ref{CaseStudy} applies CEKER to unikernel security, demonstrating its utility through a case study. Section \ref{Discussion} discusses the broader implications of the approach and its limitations. Section \ref{FW} provides potential avenues for future work, and Section \ref{Conclusion} concludes the paper.


\section{General Approach: CEKER} \label{GeneralApproach}
The CEKER approach presented in this paper is formulated to provide researchers with a systematic, repeatable, and efficient method for corpus-based literature analysis. CEKER can be flexibly applied across a wide range of research domains, from cybersecurity and computer science to healthcare and social sciences. At its heart, CEKER leverages LLM analysis, and supports the selection of any modern LLM. Each step in the approach is described in detail below.

\subsection{Step 1: Collect corpus}
The first step in the approach is to establish a clear, well-defined corpus of relevant literature for subsequent analysis. This step is intentionally flexible to support two possible scenarios: 

\begin{enumerate}
\item \textit{Reuse an existing corpus}: In some research contexts, a previously constructed set of literature may already exist. Reusing such a collection saves time and effort while ensuring consistency. Researchers may select prior literature from past systematic reviews, prior research studies, or internal datasets.
\item \textit{Collect new literature}: If no suitable corpus exists, researchers can collect a new set. This can be done using traditional academic search engines or modern AI-based tools that support intelligent search and filtering. Search queries can be tailored to necessarily limit results to relevant materials. For example, speaking hypothetically, an LLM could be broadly asked, ``Provide 25 highly cited research articles that link blue orangutans to urban development." The results would then be checked for accuracy and added to the corpus for further evaluation. It is possible for an LLM to generate non-existent papers (i.e., LLM text generation may creatively constructs ``new" papers), and this tendency may increase as more papers are requested. Before adding any papers to the corpus, they should be vetted for availability and relevance. This process can repeat until the corpus reaches a point of diminishing returns (i.e., very few new papers are being found) or the size of the corpus is sufficiently large. 
\end{enumerate}

Note that these two approaches are not mutually exclusive and can be combined as needed (i.e., an existing corpus might be augmented by AI search results). Members of this set might be further vetted (by manual inspection or AI screening) to further refine the results.

\subsection{Step 2: Extract key features}
The next step extracts critical insights from the collected literature using LLMs. LLMs offer significant advantages over traditional keyword-based approaches, as they can understand context (on a grander scale), identify implicit themes, and generate qualitative insights.

\begin{enumerate}
\item \textit{Prompt design}: To effectively extract relevant features, identify key themes, and highlight knowledge gaps, carefully designed prompts guide the LLMs. For example, prompts may include:
  \begin{itemize}
  \item ``Identify and summarize key points discussed in this paper."
  \end{itemize}
  The prompts can be adjusted based on the themes of interest. Further, prompts can be included to check for hallucinations.
\item \textit{Extraction of features and trends}: LLMs process each paper in the corpus, extracting important features, research methods, findings, and areas requiring further exploration. 
\end{enumerate}

Results are aggregated from this process for analysis in the next step. Papers that don't sufficiently align with the research topic can be filtered out.

\subsection{Step 3: Analyze and evaluate results}

The final step of the approach uses the LLM to analyze and evaluate the extracted data. Through careful prompting, the LLM can produce a variety of analyses. For example, the following prompt might be used to uncover trends:

\begin{itemize}
  \item ``List the most frequently discussed trends in this literature."
  \end{itemize}
Additional analysis by the LLM might include the following:
\begin{enumerate}
\item \textit{Thematic analysis}: LLMs categorize extracted insights into themes and topics, identifying patterns, relationships, and outliers.
\item \textit{Gap analysis}: LLMs identify knowledge gaps by pinpointing underrepresented themes and areas with limited research coverage. This allows researchers to suggest directions for future studies.
\item \textit{Visualization and reporting}: Depending on LLM capabilities, data visualizations, such as bar charts, heat maps, or word clouds, might be created to visually represent trends in the literature. 
\item \textit{Comparison to prior research}: If applicable, the results from this process can be compared against prior studies to highlight new findings or confirm previously established insights.
\end{enumerate}
Prompts should be adjusted to give the results desired. Invalid results might suggest prompts that need to be adjusted, so prelimiary results from new prompts should be quality checked.
 
Collectively, the three steps in CEKER present a thorough overview of trends, gaps, and emerging themes within a given research domain. This general approach not only aims to enhance the rigor of systematic reviews but can also accelerate the process of identifying key areas for future research.

Note that for LLMs with a larger context window, steps 2 and 3 might be combined by feeding all papers from Step 1 into the LLM and asking it the summary questions from Step 3. For example, the text could be extracted from each paper and then added to one combined text file with the papers separated by delimiters. The LLM could then analyze all of the text at once. Eventually, LLMs may support uploading a zip file of pdf documents for analysis.


\section{Case Study: Application to Unikernel Security} \label{CaseStudy}
By following the systematic three-step methodology in CEKER, we demonstrate how the general approach is applied to the specific domain of unikernel security. This case study not only validates the effectiveness of CEKER but also highlights its ability to generate new actionable insights into security features, trends, and knowledge gaps within unikernel research. ChatGPT-4o~\cite{chatgpt} was selected as the LLM for this case study, but another modern LLM could be chosen. The focus of this research is not in ``proving'' which LLM is ``superior'' for the CEKER approach, but rather that CEKER can be capably applied.

\subsection{Unikernels} \label{Unikernels}
Unikernels are an evolution of an old idea: Library Operating Systems (LibOS). The principle idea behind LibOSs is that OS functionality is implemented in the form of libraries, customized to increase performance of the application by providing closer access to hardware \cite{RethinkingLibOS}. Utilization of libraries, in addition to a unified address space, enables replacing syscalls and context switches with simpler function calls as well \cite{Unikernel:LibOSForTheCloud, RethinkingLibOS, unikernelIsolation_MPK}. To help achieve the reduced size, many familiar security features have also been removed including Address Space Layout Randomization (ASLR), Non-executable Bits (NX bits), and stack canaries \cite{michaels2019assessing, Unikernel:LibOSForTheCloud, UnikernelSecurityPerspective}. Adhering to the reductionist principle adopted by unikernels, these features, in addition to others, have been removed to save space, simplify the code base, and eliminate extra, unnecessary features \cite{Unikernel:LibOSForTheCloud, michaels2019assessing, UnikernelSecurityPerspective}.

The topic of unikernel security was selected for multiple reasons. Existing research \cite{wollman2024survey, michaels2019assessing, UnikernelSecurityPerspective} indicates that there are potential security challenges facing unikernels, possibly due to removing security features, so it represents a research area where contributions could prove valuable. Except for one paper \cite{wollman2024survey}, researchers were unable to identify other papers that conducted a literature review focused on unikernel security. Furthermore, an existing unikernel study \cite{wollman2024survey} has provided initial quantitative results with which to compare CEKER's results. Thus this domain offers an opportunity for CEKER to provide valuable insights.

\subsection{Step 1: Collect corpus}
The first step of CEKER collects a corpus of literature for analysis in the later stages. For the topic of unikernel security, a previously collected corpus of 33 vetted unikernel security papers, filtered from an initial collection of 51 papers and detailed in \cite{wollman2024survey}, is used in this case study. 
Had no papers been available, or if additional papers were desired, this stage would involve an additional literature search, either traditional or AI assisted. Table \ref{uni-search} shows the results of prompting ChatGPT-4o, ``Find 25 highly cited research articles that focus on unikernel security." Vetting these papers reveals a number of non-existent ones, with the propensity of invalid titles increasing as the size of the list (requested in one batch) increases. The papers related to the valid titles would be sourced for addition to the corpus. 

\begin{table}[h!]
\centering
\caption{Unikernel Literature Search Results (25) with ChatGPT-4o. Nonexistent papers are crossed out.}
\label{uni-search}
\begin{tabular}{|>{\setlength\parindent{0em}\setlength\hangindent{1.5em}\setlength\hangafter{1}}p{8.2cm}|}
\hline
Title \\ \hline
1. A Security Perspective on Unikernels \\
2. A Survey of Unikernel Security: Insights and Trends from a Quantitative Analysis \\
3. Unikernels Motivations, Benefits and Issues: A Multivocal Literature Review \\
4. Enhancing Cloud Security and Privacy: The Unikernel Solution \\
5. Unikernel-based Approach for Software-Defined Security in Cloud Infrastructures \\
6. Unikernel Linux (UKL) \\
7. Are Unikernels Ready for Serverless on the Edge? \\
8. Isolating Real-Time Safety-Critical Embedded Systems via SGX-based Lightweight Virtualization \\
9. The Endokernel: Fast, Secure, and Programmable Subprocess Virtualization \\
10. Unikernels: Library Operating Systems for the Cloud \\
11. IncludeOS: A Minimal, Resource Efficient Unikernel for Cloud Services \\
12. HermitCore: A Unikernel for Extreme Scale Computing \\
13. ClickOS and the Art of Network Function Virtualization \\
14. OSv—Optimizing the Operating System for Virtual Machines \\
15. \st{Rumprun: Efficient and Secure Application Execution with Unikernels} \\
16. Unikraft: Fast, Specialized Unikernels the Easy Way \\
17. Graphene-SGX: A Practical Library OS for Unmodified Applications on SGX \\
18. \st{Security Isolation of Unikernels on Xen Hypervisor} \\
19. \st{Unikernelization of Containerized Applications to Reduce Attack Surfaces} \\
20. \st{Tackling Cloud Security with Unikernels: Architectural Paradigm for Trustworthy Cloud Workloads} \\
21. \st{Sledge: A Secure, Efficient Edge Computing System Built on Unikernels} \\
22. \st{Reducing Security Complexity in IoT Devices Using Unikernels} \\
23. \st{Microsecond-Scale Isolation for Unikernels} \\ 
24. \st{Securing Smart Contracts Using Unikernel Technology} \\
25. \st{Unikernel Security: Design and Mitigation of Attack Vectors in Minimal OS Environments} \\ \hline
 \end{tabular}
\end{table}

\subsection{Step 2: Extract Key Features}
The second step of CEKER 
extracts key features from each paper. Table \ref{ChatGPTInitialPrompts} contains the unikernel prompts tailored for this phase. As discussed in \ref{Unikernels}, a need exists to better understand the state of unikernel security research. Previous research was able to generate quantitative values for specific security terms, however the breadth of security terms utilized was limited \cite{wollman2024survey}. \textbf{P-1} was crafted specifically to address this limitation and collect all the security features discussed within each paper. \textbf{P-2} uses this list and generates a quantitative score, demonstrating its relevance to the paper (and unikernel.) The purpose is to account for the context of security terms and the variability security terms can have.

\begin{table}[htbp]
    \caption{ChatGPT-4o Unikernel Security -- Prompts for CEKER Step 2 (Extract Key Features)}
    \begin{tabular}{|m{29em}|}  
        \hline
        \textbf{P-1:} What are the security features discussed in the provided paper and how do they relate to the unikernels discussed in the paper?\\
        \\
        \textbf{P-2:} Using the provided paper and the security features obtained above, provide a score between 0.00-1.00 to two decimal places that describes their relevance to the paper and the unikernel discussed in the paper.\\
        \\
        \textbf{P-3:} Using the provided paper, for each of the security terms (ASLR, DEP, Stack Canaries), provide a score between 0.00-1.00 to two decimal places that describes their relevance to unikernels discussed in the paper.\\
        \\
        \textbf{P-4:} Using the provided paper, for each of the security terms (ASLR, DEP, Stack Canaries), provide a score between 0.00-1.00 to two decimal places that describes their relevance to the paper. Also provide your confidence in the score (low, medium, high) along with an explanation for the score.\\
        \\
        \textbf{P-5:} How was the relevance score calculated for the above question?\\
        \\
        \textbf{P-6:} What are the unikernels discussed in the provided paper and what security features are discussed in relation to the unikernels discussed in the provided paper?\\
        \hline
    \end{tabular}
    \label{ChatGPTInitialPrompts}
\end{table}

\textbf{P-3} and \textbf{P-4} were crafted to collect data on specific security terms that research indicated had been removed for various reasons \cite{Unikernel:LibOSForTheCloud, michaels2019assessing, UnikernelSecurityPerspective}. The additional analysis in Step 3 of these results will provide further insight into the potential impact of these features being removed.

\textbf{P-5} was an additional check against hallucinations, providing the ability to verify results.

\textbf{P-6} states ``unikernels'' first and then ``security features.'' This is in contrast to the wording of \textbf{P-2} and \textbf{P-3} which both state the term ``security features'' first and then ``unikernels.'' The intention is to invert the priority of the terms in this prompt and evaluate if this impacts the results. Additionally, this data gives additional context to the security features discovered in earlier prompts.

Each prompt was crafted with intentionally redundant language to prevent hallucinations and ambiguity. Many of the prompts use the language ``...discussed in the paper" to conclude a sentence that started with ``Using the provided paper...". Wrapping the prompt in this way reduces the possibility ChatGPT uses data external to the paper.

The data generated from each prompt was saved as a separate document for use in Step 3. Each of the initial six prompts had 33 total responses, one for each paper in the corpus, resulting in 198 total responses. Raw results are not shown here due to space limitations. The word count for each response is shown in Table \ref{ResponseInfo}.

\begin{table}[htbp]
    \caption{Word Count of the Unikernel Security Documents Produced for each Prompt in CEKER Step 2 (Extract Key Features)}
    \centering
    \begin{tabular}{|c|c|}  
        \hline
        Prompt & Total Words \\ \hline
        P-1 & 12,204 \\ \hline
        P-2 & 10,252 \\ \hline
        P-3 & 9,297 \\ \hline
        P-4 & 10,691 \\ \hline
        P-5 & 14,873 \\ \hline
        P-6 & 16,293 \\ \hline
    \end{tabular}
    \label{ResponseInfo}
    \end{table}

\subsection{Step 3: Analyze and Evaluate Results}

The third step of CEKER uses the LLM to analyze and evaluate the extracted data to produce summary insights. Tables \ref{GeneralPrompts} and \ref{SpecificPrompts} contain the unikernel prompts tailored for this phase. Each of the six documents from Step 2 were uploaded to new ChatGPT sessions. Separate sessions were employed to avoid cross-contamination of contexts and results. Each document was asked the prompts in Table \ref{GeneralPrompts} and then the prompt-specific prompts in Table \ref{SpecificPrompts} were asked to the appropriate prompt: \textbf{SP-1} and \textbf{SP-2} asked to \textbf{P-1}, etc. The only prompt which did not receive follow-on analysis was \textbf{P-5}.

\begin{table}[htbp]
    \caption{ChatGPT-4o Unikernel Security -- General Prompts for CEKER Step 3 (Analyze and Evaluate)}
    \begin{tabular}{|m{29em}|}  
        \hline
        \textbf{GP-1:} The provided document contains 33 responses, one response per unique research paper. Summarize all the responses.\\
        \hline\\
        \textbf{GP-2:} Using the above summaries, what conclusions can you draw?\\
        \hline
    \end{tabular}
    \label{GeneralPrompts}
    \end{table}

The first prompt in Table \ref{GeneralPrompts} was created to identify and extract similarities from within each of the documents. Similar in principle to \textbf{P-1}, the vague command to "summarize" is intentional to identify themes. The second prompt intends to draw out patterns and generate an overall picture of the document; distilling the prompt to its key discussion points. The prompts in Table \ref{SpecificPrompts} were created to provide an extra level of insight specific to the original prompts (Table \ref{ChatGPTInitialPrompts}.)

\begin{table}[htbp]
    \caption{ChatGPT-4o Unikernel Security -- Specific Prompts for CEKER Step 3 (Analyze and Evaluate)}
    \begin{tabular}{|m{29em}|}  
        \hline
        \textbf{Prompt 1} \\
        \hline
        \textbf{SP-1:} Using the provided document, what is the most common security feature discussed in the summaries, and what is the most frequently identified gap in the summaries?\\
        \textbf{SP-2:} Using the provided document, what are the top three unikernels discussed the most frequently?\\
        \hline
        \textbf{Prompt 2} \\
        \hline
        \textbf{SP-3:} The provided document contains 33 responses, one response per unique research paper, scoring between 0.00-1.00 to two decimal places, the relevance security features has to the paper. What is the most common security feature, and what does the response say about its context as it relates to the paper?\\
        \textbf{SP-4:} The provided document contains 33 responses, one response per unique research paper, scoring between 0.00-1.00 to two decimal places, the relevance security features has to the paper. What is the least common security feature, and what does the response say about its context as it relates to the paper?\\
        \hline
        \textbf{Prompt 3} \\
        \hline
        \textbf{SP-5:} The provided paper has 33 responses, one response per unique research paper, to the question “Using the provided paper, for each of the security terms (ASLR, DEP, Stack Canaries), provide a score between 0.00-1.00 to two decimal places that describes their relevance to unikernels discussed in the paper.” Organize the papers from the scores closest to 0.00 to the scores closest to 1.00\\
        \textbf{SP-6:} Using the provided document, which paper has the highest score for ASLR? Which paper has the highest score for DEP? Which paper has the highest score for Stack Canaries? Which paper has the lowest score for ASLR? Which Paper has the lowest score for DEP? Which paper has the lowest score for Stack Canaries?\\
        \hline
        \textbf{Prompt 4} \\
        \hline
        \textbf{SP-7:} The provided paper has 33 responses, one response per unique research paper, to the question “Using the provided paper, for each of the security terms (ASLR, DEP, Stack Canaries), provide a score between 0.00-1.00 to two decimal places that describes their relevance to the paper. Also provide your confidence in the score (low, medium, high) along with an explanation for the score.” Organize the papers from highest to lowest confidence score\\
        \hline
        \textbf{Prompt 6} \\
        \hline
        \textbf{SP-8:} The provided paper has 33 responses, one response per unique research paper, to the question “What are the unikernels discussed in the provided paper and what security features are discussed in relation to the unikernels discussed in the provided paper?” Identify what is the most commonly discussed unikernel and what its security features are.\\
        \textbf{SP-9:} Using the provided document, what are the top five most commonly discussed unikernels and the top five most commonly discussed security features\\
        \hline
    \end{tabular}
    \label{SpecificPrompts}
\end{table}

After the analysis was completed, the following results were observed.

\subsubsection{Themes} \label{Results_Themes}
Within this corpus six themes were identified, as shown in Table \ref{IdentifiedThemes}. Of the six, the three themes in bold were further identified as the most significant; appearing in nearly every paper. Reduced attack surface was singled out further and identified as the most common feature. Minimalism and a single address space were frequently cited as elements contributing to the reduced attack surface theme. It also frequently scored close to, or at, 1.00 on relevance scores. The themes sometimes shared overlapping reasons for their ranking. While single address space was attributed to reduced attack surface, it was also attributed to isolation mechanisms and customization. Similarly, minimalism was a key component to reduced attack surface and performance-security.

The Advanced Security Features theme contained a variety of topics, both positive and negative. On the positive, hardware specific security components were identified as significant elements to this theme: Software-guard Extensions (SGX), Memory Protection Keys (MPK), memory sealing and memory isolation. On the negative, missing stack protections, debugging tools, and entropy generation were listed. Though not specifically identified as a theme, missing stack protections and entropy generation frequently appeared as problem areas for unikernels. This will be discussed further in \ref{Results_Gaps}.

The Customization and Adaptability theme contained an interesting set of topics. Internet of Things (IoT), serverless computing, and cloud environments were large influences on this theme's significance. Two research efforts were even suggested in alignment with this theme; more cloud-native orchestration tools and a hybrid container-unikernel environment.

\begin{table}[htbp]
    \caption{Identified Themes (Unikernel Security) -- CEKER Step 3 Results}
    \centering
    \begin{tabular}{|c|}  
        \hline
        \textbf{Reduced Attack Surface}\\
        \hline
        \textbf{Isolation}\\
        \hline
        \textbf{Immutable infrastructure}\\
        \hline
        Performance-security balance\\
        \hline
        Customization and Adaptability\\
        \hline
        Advanced Security Features\\
        \hline
    \end{tabular}
    \label{IdentifiedThemes}
\end{table}

\subsubsection{Gaps} \label{Results_Gaps}
There were several identified gaps in the corpus, shown in Table \ref{IdentifiedGaps}. The biggest gap was Missing Traditional OS Features. Dynamic Security Adjustments were further identified as the least common security feature in the corpus. This feature would enable on-the-fly regeneration of unikernels and enable unikernels to adapt to real-time threats. ASLR was also frequently discussed as a gap, however there were differing ideas as-to its significance. Many of the research papers focused on ASLR as a solution, however some implementations prefer a dynamic form such as function-based ASLR to by-pass the limitations imposed by the single address space. Entropy generation was another discussed gap, attributed to limited entropy sources and resource limited environments. 

Three security features were specified in the prompts: ASLR, DEP, stack canaries. ASLR was the most common term discussed in the corpus, as discussed above. The specific term DEP did not appear frequently but was generally associated with the idea of privilege separation, which was prevalent in the corpus. Of the three, stack canaries appeared least frequently and was often disregarded as a security feature. The reasons for this often came back to compile-time guarantees and the focus on unikernel simplicity.

\begin{table}[htbp]
    \caption{Identified Gaps (Unikernel Security) -- CEKER Step 3 Results}
    \centering
    \begin{tabular}{|c|}  
        \hline
        \textbf{Missing Traditional OS Features}\\
        \hline
        Difficulty Debugging\\
        \hline
        Dependency on Hypervisors\\
        \hline
    \end{tabular}
    \label{IdentifiedGaps}
\end{table}

\subsubsection{Comparisons} \label{Results_Comparisons}
\begin{figure}[h]
\centering
\includegraphics[scale=.34]{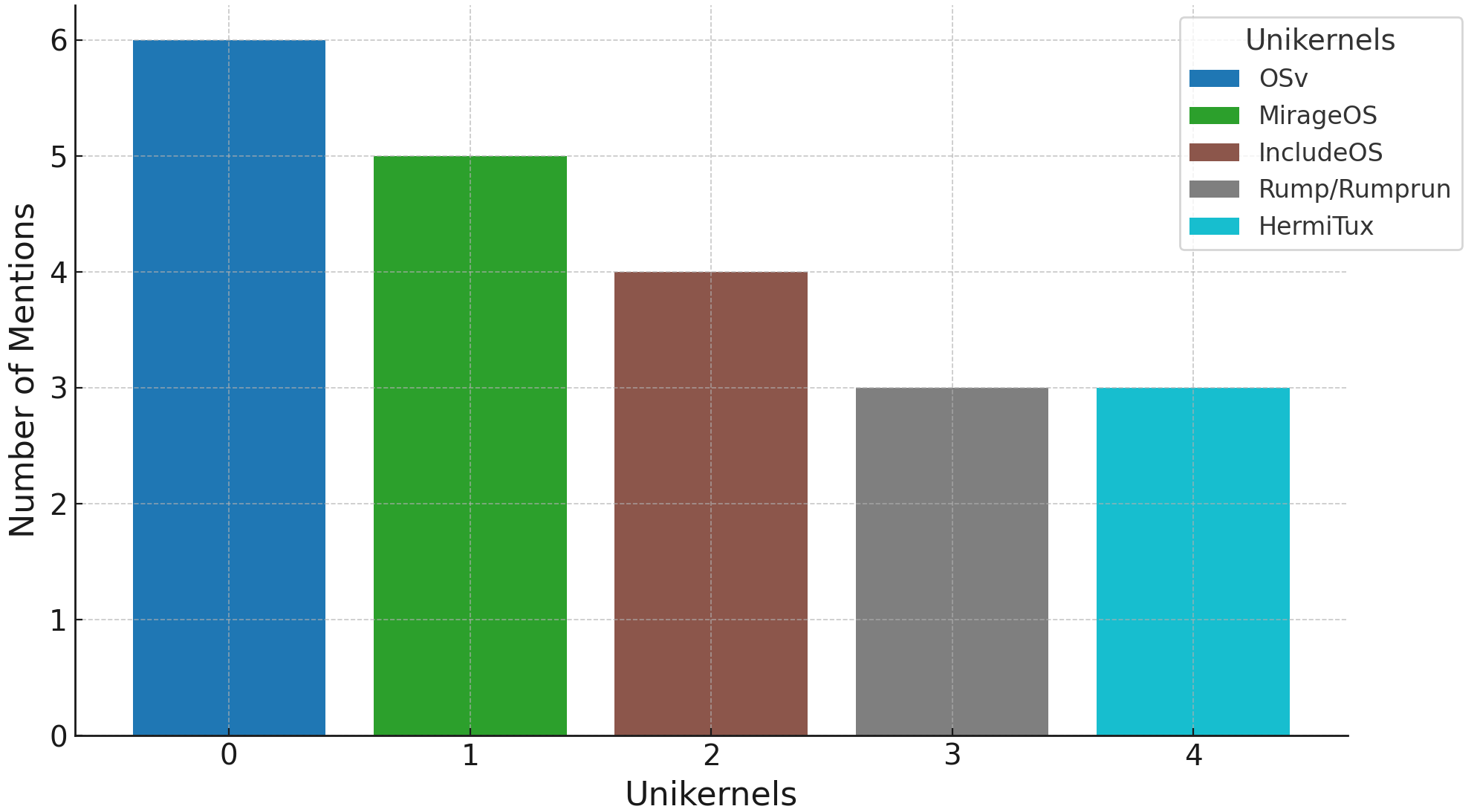}
\caption{Common Unikernels in Unikernel Research}
\label{Fig_Unikernel}
\end{figure}

Figure \ref{Fig_Unikernel} shows the number of mentions for the most commonly discussed unikernels as generated by ChatGPT-4o. The calculation for number of mentions was based on direct mentions, contextual significance, and appearance across papers. Comparing these results to previous research the findings are similar, with HermiTux included in CEKER's top 5 instead of Graphene-SGX or Ling \cite{wollman2024survey}.

\begin{figure}[h]
\centering
\includegraphics[scale=.34]{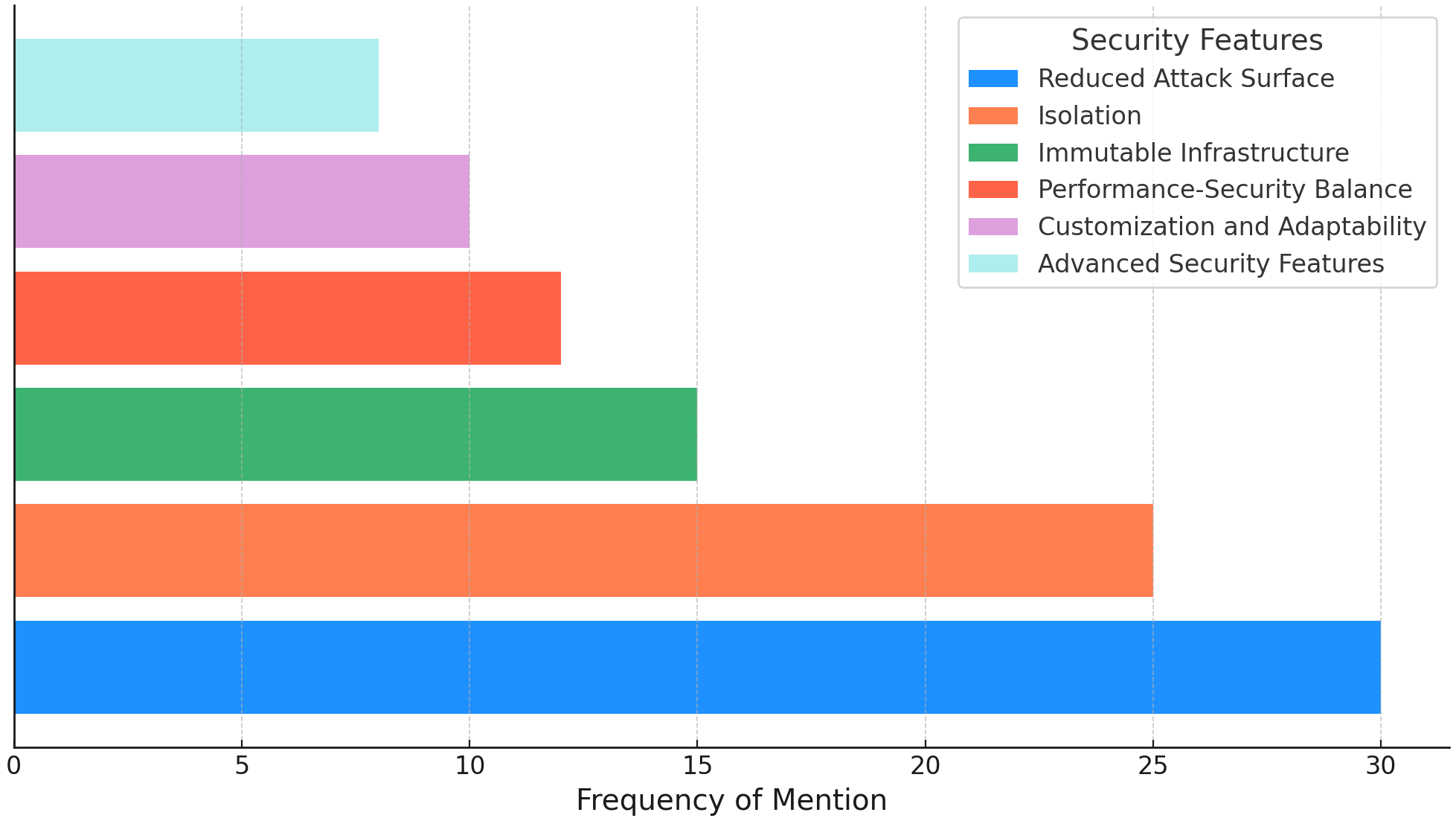}
\caption{Common Security Features in Unikernel Research}
\label{Fig_Security}
\end{figure}

The data shown in Figure \ref{Fig_Security} shows the frequency of mentions for the most common security features as generated by ChatGPT-4o. The calculation for these values were based on document review, feature tally, and aggregation. In this instance feature tally was impacted by the significance of the term in the paper. Comparing these results to previous research \cite{wollman2024survey} reveals stark differences; there is no overlap. The results produced by CEKER however tell a more complete story. Specific security terms can be generalized into themes that then appear more frequently throughout the literature, compared to looking specifically for ASLR.

\section{Discussion} \label{Discussion}

Because of the flexibility of CEKER, applying it to unikernel security research was simple and intuitive. Ensuring the prompts for Step 2 returned accurate information required some consideration and experimentation, but not overly burdensome. Generating two sets of prompts for Step 3 resulted in more impactful data than expected. \textbf{GP-1} generated very thorough summaries comprised of three to four lists consisting of six to seven elements all containing multiple sentence descriptions. Human analysis of the results, combined with \textbf{GP-2} data, generated a concise and accurate portrait of the corpus. Based on the researchers' subject expertise, reviewing the results did not reveal hallucinations nor false or inaccurate information. 

The biggest strength of CEKER is demonstrated through its theme and gap generation. LLMs can process vast amounts of data and generate detailed summaries such that the human counterpart would require weeks and multiple readings of papers to achieve, and only achieve the theme. Gap identification, if not explicitly identified in research, is an extremely challenging component. LLMs once again can process the data and provide, at minimum, missing information it identified. In this case study it identified traditional OS features as a gap in unikernel security research. This is indeed a specifically noted gap in many research papers, but also alluded to through discussion of omitted security features (e.g., ASLR, memory protections).

Noted limitations in \cite{wollman2024survey} were the inability to account for all possible variations of a security feature's name and account for variations in context of a security feature within a paper. The CEKER process enabled prompt generation to account for variations in names in a simple fashion (\textbf{P-1} in Table \ref{ChatGPTInitialPrompts} for example). A programmatic approach could not exhaustively account for all possibilities, and a manual approach would be infeasible given a large enough corpus, or a varied enough set of search elements.

A notable finding in this case study was the two research efforts ChatGPT offered: cloud native orchestration tools and a container-unikernel hybrid. There are research papers in the corpus which discuss cloud integrations of unikernels \cite{CloudCyberSecurity, OnTheFlyUnikernel, TOSCABasedUnikernelApproach, UnikernelApproachCloud}, and also papers which compare and contrast different virtualization technologies \cite{VirtConTax, CompVirtandCont}. This demonstrates how CEKER can not only be used in analysis, but also the generation of research ideas from the corpus. CEKER can begin anew with this seed idea and re-evaluate at each step of CEKER as necessary.

\section{Future Work} \label{FW}
In this research the only LLM utilized was ChatGPT-4o. Other LLMs have different strengths which can be utilized to generate different results. For example Google's Gemini has a larger context window, which could perhaps enable it to understand data in a larger document. This would be beneficial, as noted in Section \ref{GeneralApproach} to provide the LLM with one single document containing all the research papers instead of uploading individual ones. Gemini's ability in this area remains to be tested.

Many LLMs also provide API access, which enables programmatic solutions and further decreases the time needed for CEKER. While CEKER is able to utilize API access, this research relied on direct interaction with the ChatGPT website in order to demonstrate proof of concept. Automation would also simplify utilizing additional LLMs (after the initial programming effort) as prompts could be submitted simultaneously, or directed by user interactions.

\section{Conclusion} \label{Conclusion}
Utilizing LLMs to supplement, or replace, time-consuming steps in the literature analysis process,  this research introduced a novel three-step approach called CEKER: \textbf{C}ollect corpus, \textbf{E}xtract \textbf{K}ey features, Analyze and \textbf{E}valuate \textbf{R}esults. CEKER can be conveniently applied to any research domain due to its flexible and general approach utilizing LLMs. 

A case study on unikernel security illustrates CEKER's ability to generate novel insights validated against previous manual methods. CEKER’s analysis highlighted reduced attack surface as the most prominent unikernel security theme. Key security gaps included the absence of ASLR, missing debugging tools, and limited entropy generation, all of which represent important challenges to unikernel security. The study also revealed a reliance on hypervisors as a potential attack vector and emphasized the need for dynamic security adjustments to address real-time threats. As seen through its application in the case study, CEKER's rapid, repeatable process is capable of uncovering research avenues, thereby supporting a continuous research process.

\balance
\printbibliography

@article{Unikernel:LibOSForTheCloud,
author = {Madhavapeddy, Anil and Mortier, Richard and Rotsos, Charalampos and Scott, David and Singh, Balraj and Gazagnaire, Thomas and Smith, Steven and Hand, Steven and Crowcroft, Jon},
title = {Unikernels: Library Operating Systems for the Cloud},
year = {2013},
issue_date = {March 2013},
publisher = {Association for Computing Machinery},
address = {New York, NY, USA},
volume = {41},
number = {1},
%issn = {0163-5964},
%note={\url{https://doi.org/10.1145/2490301.2451167}},
%url = {https://doi.org/10.1145/2490301.2451167},
doi = {10.1145/2490301.2451167},
abstract = {We present unikernels, a new approach to deploying cloud services via applications written in high-level source code. Unikernels are single-purpose appliances that are compile-time specialised into standalone kernels, and sealed against modification when deployed to a cloud platform. In return they offer significant reduction in image sizes, improved efficiency and security, and should reduce operational costs. Our Mirage prototype compiles OCaml code into unikernels that run on commodity clouds and offer an order of magnitude reduction in code size without significant performance penalty. The architecture combines static type-safety with a single address-space layout that can be made immutable via a hypervisor extension. Mirage contributes a suite of type-safe protocol libraries, and our results demonstrate that the hypervisor is a platform that overcomes the hardware compatibility issues that have made past library operating systems impractical to deploy in the real-world.},
journal = {SIGARCH Comput. Archit. News},
month = {3},
pages = {461–472},
numpages = {12},
keywords = {hypervisor, microkernel, functional programming}
}

@inproceedings{RethinkingLibOS,
author = {Porter, Donald E. and Boyd-Wickizer, Silas and Howell, Jon and Olinsky, Reuben and Hunt, Galen C.},
title = {Rethinking the Library OS from the Top Down},
year = {2011},
%isbn = {9781450302661},
publisher = {Association for Computing Machinery},
address = {New York, NY, USA},
%note = {\url{https://doi.org/10.1145/1950365.1950399}},
%url = {https://doi.org/10.1145/1950365.1950399},
doi = {10.1145/1950365.1950399},
abstract = {This paper revisits an old approach to operating system construc-tion, the library OS, in a new context. The idea of the library OS is that the personality of the OS on which an application depends runs in the address space of the application. A small, fixed set of abstractions connects the library OS to the host OS kernel, offering the promise of better system security and more rapid independent evolution of OS components.We describe a working prototype of a Windows 7 library OS that runs the latest releases of major applications such as Microsoft Excel, PowerPoint, and Internet Explorer. We demonstrate that desktop sharing across independent, securely isolated, library OS instances can be achieved through the pragmatic reuse of net-working protocols. Each instance has significantly lower overhead than a full VM bundled with an application: a typical application adds just 16MB of working set and 64MB of disk footprint. We contribute a new ABI below the library OS that enables application mobility. We also show that our library OS can address many of the current uses of hardware virtual machines at a fraction of the overheads. This paper describes the first working prototype of a full commercial OS redesigned as a library OS capable of running significant applications. Our experience shows that the long-promised benefits of the library OS approach better protection of system integrity and rapid system evolution are readily obtainable.},
booktitle = {Proceedings of the Sixteenth International Conference on Architectural Support for Programming Languages and Operating Systems},
month = {3},
pages = {291–304},
numpages = {14},
keywords = {libos, drawbridge, library os},
location = {Newport Beach, California, USA},
series = {ASPLOS XVI}
}

@inproceedings{CompVirtandCont,
  title={Virtualization and containerization of application infrastructure: A comparison},
  author={Scheepers, Mathijs Jeroen},
  booktitle={21st twente student conference on IT},
  volume={21},
  pages={1--7},
  year={2014},
  %note={\url{https://thijs.ai/papers/scheepers-virtualization-containerization.pdf} Accessed 1/16/2024},
  %url={https://thijs.ai/papers/scheepers-virtualization-containerization.pdf},
url={https://web.archive.org/web/20240928043944/https://thijs.ai/papers/scheepers-virtualization-containerization.pdf},
urldate={2024-12-03}
}

@incollection{CloudCyberSecurity,
author = {Bob Duncan and Andreas Happe and Alfred Bratterud},
title = {Cloud Cyber Security: Finding an Effective Approach with Unikernels},
booktitle = {Advances in Security in Computing and Communications},
publisher = {IntechOpen},
address = {Rijeka},
year = {2017},
editor = {Jaydip Sen},
chapter = {2},
doi = {10.5772/67801},
%note = {\url{https://doi.org/10.5772/67801}},
%url = {https://doi.org/10.5772/67801}
}

@INPROCEEDINGS{UnikernelApproachCloud,
  author={Compastié, Maxime and Badonnel, Rémi and Festor, Olivier and He, Ruan and Kassi-Lahlou, Mohamed},
  booktitle={NOMS 2018 - 2018 IEEE/IFIP Network Operations and Management Symposium}, 
  title={Unikernel-based approach for software-defined security in cloud infrastructures}, 
  year={2018},
  pages={1-7},
%  note={\url{https://doi.org/10.1109/NOMS.2018.8406155}},
  doi={10.1109/NOMS.2018.8406155}
}

@INPROCEEDINGS{UnikernelSecurityPerspective,
  author={Talbot, Joshua and Pikula, Przemek and Sweetmore, Craig and Rowe, Samuel and Hindy, Hanan and Tachtatzis, Christos and Atkinson, Robert and Bellekens, Xavier},
  booktitle={2020 International Conference on Cyber Security and Protection of Digital Services (Cyber Security)}, 
  title={A Security Perspective on Unikernels}, 
  year={2020},
  pages={1-7},
  %note={\url{https://doi.org/10.1109/CyberSecurity49315.2020.9138883}},
  doi={10.1109/CyberSecurity49315.2020.9138883}
}

@techreport{michaels2019assessing,
  title={Assessing unikernel security},
  institution={NCC Group},
  author={Michaels, Spencer and Dileo, Jeff},
  year={2019},
urldate={2024-12-03},
  %note={\url{https://research.nccgroup.com/2019/02/04/assessing-unikernel-security/} Accessed 1/16/2024},
  url={https://people.cs.pitt.edu/~babay/courses/cs3551/projects/SCADA_Unikernel/NCC_Group-Assessing_Unikernel_Security.pdf}
}

@inproceedings{unikernelIsolation_MPK,
author = {Sung, Mincheol and Olivier, Pierre and Lankes, Stefan and Ravindran, Binoy},
title = {Intra-Unikernel Isolation with Intel Memory Protection Keys},
year = {2020},
month = {3},
%isbn = {9781450375542},
publisher = {Association for Computing Machinery},
address = {New York, NY, USA},
%note = {\url{https://doi.org/10.1145/3381052.3381326}},
%url = {https://doi.org/10.1145/3381052.3381326},
doi = {10.1145/3381052.3381326},
abstract = {Unikernels are minimal, single-purpose virtual machines. This new operating system model promises numerous benefits within many application domains in terms of lightweightness, performance, and security. Although the isolation between unikernels is generally recognized as strong, there is no isolation within a unikernel itself. This is due to the use of a single, unprotected address space, a basic principle of unikernels that provide their lightweightness and performance benefits. In this paper, we propose a new design that brings memory isolation inside a unikernel instance while keeping a single address space. We leverage Intel's Memory Protection Key to do so without impacting the lightweightness and performance benefits of unikernels. We implement our isolation scheme within an existing unikernel written in Rust and use it to provide isolation between trusted and untrusted components: we isolate (1) safe kernel code from unsafe kernel code and (2) kernel code from user code. Evaluation shows that our system provides such isolation with very low performance overhead. Notably, the unikernel with our isolation exhibits only 0.6\% slowdown on a set of macro-benchmarks.},
booktitle = {Proceedings of the 16th ACM SIGPLAN/SIGOPS International Conference on Virtual Execution Environments},
pages = {143–156},
numpages = {14},
keywords = {memory protection keys, memory safety, unikernels},
location = {Lausanne, Switzerland},
series = {VEE '20}
}

@INPROCEEDINGS{VirtConTax,
  author={Sahoo, Jyotiprakash and Mohapatra, Subasish and Lath, Radha},
  booktitle={2010 Second International Conference on Computer and Network Technology}, 
  title={Virtualization: A Survey on Concepts, Taxonomy and Associated Security Issues}, 
  year={2010},
  volume={},
  number={},
  pages={222-226},
  doi={10.1109/ICCNT.2010.49},
%  url={https://doi.org/10.1109/ICCNT.2010.49}
  }

@INPROCEEDINGS{OnTheFlyUnikernel,
author={Compastié, Maxime and Badonnel, Rémi and Festor, Olivier and He, Ruan},
booktitle={NOMS 2018 - 2018 IEEE/IFIP Network Operations and Management Symposium}, 
title={Demo: On-the-fly generation of unikernels for software-defined security in cloud infrastructures}, 
year={2018},
volume={},
number={},
pages={1-2},
%note = {\url{https://doi.org/10.1109/noms.2018.8406131}},
doi={10.1109/NOMS.2018.8406131}}

@INPROCEEDINGS{TOSCABasedUnikernelApproach,
author={Compastié, Maxime and Badonnel, Rémi and Festor, Olivier and He, Ruan},
booktitle={2019 IEEE Conference on Network Softwarization (NetSoft)}, 
title={A TOSCA-Oriented Software-Defined Security Approach for Unikernel-Based Protected Clouds}, 
year={2019},
volume={},
number={},
pages={151-159},
%note = {\url{https://doi.org/10.1109/netsoft.2019.8806623}},
doi={10.1109/NETSOFT.2019.8806623}}

@inproceedings{wollman2024survey,
title={A Survey of Unikernel Security: Insights and Trends from a Quantitative Analysis}, 
author={Alex Wollman and John Hastings},
booktitle={Proceedings of the 2024 IEEE 4\,\textsuperscript{th} Cyber Awareness and Research Symposium (CARS'24)},
year={2024},
doi={10.1109/CARS61786.2024.10778787},
%eprint={2406.01872},
%archivePrefix={arXiv},
%primaryClass={cs.CR},
month={10}
}

@article{thelwall2022scopus,
    author = {Thelwall, Mike and Sud, Pardeep},
    title = {Scopus 1900–2020: Growth in articles, abstracts, countries, fields, and journals},
    journal = {Quantitative Science Studies},
    volume = {3},
    number = {1},
    pages = {37-50},
    year = {2022},
    month = {04},
    abstract = {Scientometric research often relies on large-scale bibliometric databases of academic journal articles. Long-term and longitudinal research can be affected if the composition of a database varies over time, and text processing research can be affected if the percentage of articles with abstracts changes. This article therefore assesses changes in the magnitude of the coverage of a major citation index, Scopus, over 121 years from 1900. The results show sustained exponential growth from 1900, except for dips during both world wars, and with increased growth after 2004. Over the same period, the percentage of articles with 500+ character abstracts increased from 1\% to 95\%. The number of different journals in Scopus also increased exponentially, but slowing down from 2010, with the number of articles per journal being approximately constant until 1980, then tripling due to megajournals and online-only publishing. The breadth of Scopus, in terms of the number of narrow fields with substantial numbers of articles, simultaneously increased from one field having 1,000 articles in 1945 to 308 fields in 2020. Scopus’s international character also radically changed from 68\% of first authors from Germany and the United States in 1900 to just 17\% in 2020, with China dominating (25\%).},
    issn = {2641-3337},
    doi = {10.1162/qss_a_00177},
    %url = {https://doi.org/10.1162/qss\_a\_00177},
    %eprint = {https://direct.mit.edu/qss/article-pdf/3/1/37/2008360/qss\_a\_00177.pdf},
}

@article{haddaway2015making,
  title={Making literature reviews more reliable through application of lessons from systematic reviews},
  author={Haddaway, Neal R and Woodcock, Paul and Macura, Biljana and Collins, Alexandra},
  journal={Conservation Biology},
  volume={29},
  number={6},
  pages={1596--1605},
  year={2015},
  publisher={Wiley Online Library},
doi={10.1111/cobi.12541}
}

@article{bolanos2024artificial,
  title={Artificial intelligence for literature reviews: Opportunities and challenges},
  author={Bolanos, Francisco and Salatino, Angelo and Osborne, Francesco and Motta, Enrico},
  journal={Artificial Intelligence Review},
  volume={57},
  number={10},
  pages={259},
  year={2024},
  publisher={Springer},
doi={10.1007/s10462-024-10902-3}
}

@article{ofori2024towards,
  title={Towards the automation of systematic reviews using natural language processing, machine learning, and deep learning: a comprehensive review},
  author={Ofori-Boateng, Regina and Aceves-Martins, Magaly and Wiratunga, Nirmalie and Moreno-Garcia, Carlos Francisco},
  journal={Artificial intelligence review},
  volume={57},
  number={8},
  pages={200},
  year={2024},
  publisher={Springer},
doi={10.1007/s10462-024-10844-w}
}

@inproceedings{wittenborg2024swarm,
  title={SWARM-SLR-Streamlined Workflow Automation for Machine-Actionable Systematic Literature Reviews},
  author={Wittenborg, Tim and Karras, Oliver and Auer, S{\"o}ren},
  booktitle={International Conference on Theory and Practice of Digital Libraries},
  pages={20--40},
  year={2024},
  organization={Springer},
doi={10.1007/978-3-031-72437-4_2}
}

@misc{ali2024automatedliteraturereviewusing,
      title={Automated Literature Review Using NLP Techniques and LLM-Based Retrieval-Augmented Generation}, 
      author={Nurshat Fateh Ali and Md. Mahdi Mohtasim and Shakil Mosharrof and T. Gopi Krishna},
      year={2024},
      eprint={2411.18583},
      archivePrefix={arXiv},
      primaryClass={cs.CL},
      %url={https://arxiv.org/abs/2411.18583}, 
}

@misc{joos2024cuttingclutterpotentialllms,
      title={Cutting Through the Clutter: The Potential of LLMs for Efficient Filtration in Systematic Literature Reviews}, 
      author={Lucas Joos and Daniel A. Keim and Maximilian T. Fischer},
      year={2024},
      eprint={2407.10652},
      archivePrefix={arXiv},
      primaryClass={cs.LG},
      %url={https://arxiv.org/abs/2407.10652}, 
}

@misc{chatgpt,
url={https://chatgpt.com},
urldate={2024-12-13},
title={ChatGPT 4o},
author={{OpenAI}},
year={2024}
}




\end{document}